\documentclass[times,twocolumn]{aastex631}

\usepackage[T1]{fontenc}
\usepackage{graphicx}
\usepackage{amsmath}
\usepackage{amssymb}
\usepackage{subfigure}
\usepackage{cases}
\usepackage{mathrsfs}
\usepackage[flushleft]{threeparttable}
\usepackage{amssymb}
\usepackage{pifont}
\usepackage{epstopdf}
\usepackage{xcolor}
\usepackage[all]{hypcap}
\usepackage{mwe,tikz}
\usepackage{bm}
\usepackage{multirow}
\usepackage{longtable}
\usepackage{xspace}
\usepackage{rotating}
\usepackage{CJK}
\usepackage{url}
\usepackage{upgreek}
\usepackage{booktabs}
\usepackage{textcomp}
\usepackage{makecell}
\usepackage{enumitem}
\usepackage{csquotes}
\usepackage{soul}
\usepackage{tabularx}
\usepackage{subfiles}
\usepackage{ulem}
\usepackage{hyperref}

\newcommand{\hbeta}{H{$\beta$}}
\newcommand{\MgII}{Mg\,{\sc ii}}

\newcommand{\e}{\text{e}}

\newcommand{\p}{\partial}

\journalinfo{Published in The Astrophysical Journal, 1012, 88 (2026)\quad
  \normalfont\href{https://doi.org/10.3847/1538-4357/aeb2f9}{doi:10.3847/1538-4357/aeb2f9}}
\shorttitle{EP241206a: A Sub-kilosecond X-Ray Flare}
\shortauthors{Wang et al.}

\begin{document}

\title{EP241206a: A Sub-kilosecond X-Ray Flare Possibly from an Active Galactic Nucleus with a Flux Increase over 3 Orders of Magnitude}

\correspondingauthor{Tinggui Wang, Hua Feng}
\email{twang@ustc.edu.cn; hfeng@ihep.ac.cn}

\author[0009-0007-9068-2752]{Chen-Yu Wang}
\affiliation{Department of Astronomy, Tsinghua University, Beijing 100084, China}

\author[0000-0002-1517-6792]{Tinggui Wang}
\affiliation{Department of Astronomy, School of Physical Sciences, University of Science and Technology of China}

\author[0000-0001-7584-6236]{Hua Feng}
\affiliation{State Key Laboratory of Particle Astrophysics, Institute of High Energy Physics, Chinese Academy of Sciences, 100049, Beijing, China}

\author[0000-0002-0096-3523]{Wenxiong Li}
\affiliation{National Astronomical Observatories, Chinese Academy of Sciences, 20A Datun Road, Beijing 100101, China}

\author[0009-0001-8733-2088]{Hucheng Ding}
\affiliation{Department of Physics, Anhui Normal University, Wuhu, Anhui, 241002, China}

\author[0000-0002-7020-4290]{Xinwen Shu}
\affiliation{Department of Physics, Anhui Normal University, Wuhu, Anhui, 241002, China}

\author[0000-0002-5596-5059]{Yi-Han Iris Yin}
\affiliation{Department of Physics, The University of Hong Kong, Pokfulam Road, Hong Kong, China}
\affiliation{The Hong Kong Institute for Astronomy and Astrophysics, The University of Hong Kong, Hong Kong, China}

\author[0000-0002-4622-796X]{Roberto Soria}
\affiliation{INAF—Osservatorio Astrofisico di Torino, Strada Osservatorio 20, I-10025 Pino Torinese, Italy}
\affiliation{Sydney Institute for Astronomy, School of Physics A28, The University of Sydney, Sydney NSW 2006, Australia}

\author[0000-0002-9725-2524]{Bing Zhang}
\affiliation{Department of Physics, The University of Hong Kong, Pokfulam Road, Hong Kong, China}
\affiliation{The Hong Kong Institute for Astronomy and Astrophysics, The University of Hong Kong, Hong Kong, China}

\author[0000-0001-5990-6243]{Arne Rau}
\affiliation{Max-Planck-Institut für extraterrestrische Physik, Giessenbachstraße 1, 85748 Garching, Germany}

\author[0000-0002-1935-8104]{Yongquan Xue}
\affiliation{Department of Astronomy, School of Physical Sciences, University of Science and Technology of China}

\author[0009-0009-8982-2361]{Mingjun Liu}
\affiliation{National Astronomical Observatories, Chinese Academy of Sciences, 20A Datun Road, Beijing 100101, China}

\author[0000-0002-4912-2477]{Richard Saxton}
\affiliation{Telespazio UK for ESA, ESAC, Apartado 78, 28691 Villanueva de la Cañada, Madrid, Spain}

\author[0000-0002-4419-6434]{JunXian Wang}
\affiliation{Department of Astronomy, School of Physical Sciences, University of Science and Technology of China}

\author[0000-0003-4253-656X]{D. Andrew Howell}
\affiliation{Las Cumbres Observatory, 6740 Cortona Drive Suite 102, Goleta, CA 93117-5575, USA}
\affiliation{Department of Physics, University of California, Santa Barbara, CA 93106-9530, USA}
 
\author[0000-0001-5807-7893]{Curtis McCully}
\affiliation{Las Cumbres Observatory, 6740 Cortona Drive Suite 102, Goleta, CA 93117-5575, USA}
 
\author{Joseph R. Farah}
\affiliation{Las Cumbres Observatory, 6740 Cortona Drive Suite 102, Goleta, CA 93117-5575, USA}
\affiliation{Department of Physics, University of California, Santa Barbara, CA 93106-9530, USA}
 
\author{Moira Andrews}
\affiliation{Las Cumbres Observatory, 6740 Cortona Drive Suite 102, Goleta, CA 93117-5575, USA}
\affiliation{Department of Physics, University of California, Santa Barbara, CA 93106-9530, USA}

\author[0000-0001-5249-2797]{Hai-Wu Pan}
\affiliation{National Astronomical Observatories, Chinese Academy of Sciences, 20A Datun Road, Beijing 100101, China}

\author{Donghua Zhao}
\affiliation{National Astronomical Observatories, Chinese Academy of Sciences, 20A Datun Road, Beijing 100101, China}

\begin{abstract}  
EP241206a is a fast X-ray transient detected with the Wide-field X-ray Telescope (WXT) on board the Einstein Probe. The event exhibited a single-peaked soft X-ray flare with a duration of $\sim500$ s and a peak flux of $1.0 \times 10^{-9}\ \mathrm{erg\ s^{-1}\ cm^{-2}}$ in 0.5–4 keV, spatially coincident with a spectroscopically confirmed active galactic nucleus (AGN) at $z=0.62$. Follow-up and archival observations revealed persistent X-ray emission at a level of $\sim5\times10^{-13}\ \mathrm{erg\ s^{-1}\ cm^{-2}}$, corresponding to $L_{2-10\, {\rm keV}}\simeq2.3\times10^{44}\ \mathrm{erg\ s^{-1}}$ if associated with the AGN; the WXT flare reached an estimated peak isotropic-equivalent rest-frame 2--10~keV luminosity of $\sim1.4\times10^{48}\ \mathrm{erg\ s^{-1}}$, implying a peak-to-persistent luminosity contrast of over 3 orders of magnitude. Optical spectroscopy detected broad \MgII\ and \hbeta\ emission lines, suggesting a black hole mass of $\sim3.8\times10^7 M_{\odot}$ and Eddington ratios of 0.8 at the persistent state versus an apparent peak ratio of $\sim4.6\times10^3$ during the flare. The extremely short duration and high apparent peak luminosity favor a jet origin, likely due to magnetic reconnection in the inner regions of a relativistic jet. Alternative scenarios include a
jet-in-the-jet or a gamma-ray burst occurring within the AGN disk. However, the physical association between the transient and the AGN is not firmly established; for the radio-loud and X-ray-detected AGN counterpart class, the trial-corrected chance probability is $4.9\times10^{-3}$. EP241206a may represent an example of a broader population of unidentified fast X-ray transients detected with WXT.

\end{abstract}

\keywords{X-ray transient sources (1852) --- Active galactic nuclei (16)}

\section{Introduction}

Active galactic nuclei (AGNs) are among the most energetic astrophysical systems, powered by accretion of matter onto black holes at the centers of galaxies. These systems emit across the entire electromagnetic spectrum, from radio to gamma-ray, and often exhibit variability on a wide range of timescales. One prominent manifestation of variability is the flare --- a stochastic increase in flux that can span multiple energy bands. AGN flares can be triggered by mechanisms such as instabilities in the accretion disk \citep{Czerny2009, Sniegowska2020, Raj2021}, magnetic reconnection in the corona \citep{Merloni2001, Cheng2020}, or relativistic shocks and other transient dissipation processes in the jet \citep{Marscher2008, Blandford2019, Raiteri2025}.

While many AGNs exhibit long-term variability over months to years, rapid, large-amplitude X-ray outbursts on timescales ranging from days \citep{Ulrich1997, ErinKara2025, Liu2025} down to $\sim10^3$~s \citep[e.g.,][]{Zhu2018} have also been reported in a small number of systems, including the nearby low-mass black hole in NGC~4395 \citep{Vaughan2005} and the narrow-line Seyfert-1 Mrk~766 \citep{Mochizuki2023}, with a flux increase by a factor of $\sim 5-10$. Such fast events may be associated with a variety of physical processes, including rapid changes in the inner accretion flow, coronal magnetic reconnection, or transient jet activity \citep{Merloni2001, Arvalo2006, Marscher2008, Cheng2020, Raiteri2025}. 

However, due to their rarity, short durations, and relatively low peak fluxes, our understanding of their physical origins and emission mechanisms remains limited. Wide-field X-ray survey instruments, capable of monitoring a large portion of the sky simultaneously, have thus played a crucial role in advancing the study of rapid flares \citep{Sguera2005, Yuan2022}.

The Einstein Probe (EP) mission represents a major step forward in this regard. Equipped with the Wide-field X-ray Telescope (WXT) and the Follow-up X-ray Telescope (FXT), EP is designed to perform sensitive monitoring of the soft X-ray (0.5–4 keV) sky with an unprecedented combination of wide field-of-view, fast response, and high localization accuracy \citep{Yuan2022}. Among the various transient sources detected by EP, a significant subset is suspected to originate from AGNs \citep{Liu2025}.

In this work, we focus on EP241206a, a short-duration X-ray transient potentially associated with an AGN. We analyzed multiwavelength data from radio to MeV bands to constrain the origin of this transient. In Section \ref{Data}, we describe the multiwavelength data utilized for analysis and present the observational properties of the transient. We discuss the possible origin in Section \ref{Discu}. The main conclusions are summarized in Section \ref{conclusion}.

\section{Observations and Results}\label{Data}

\subsection{X-Ray Observations and Variabilities}

On 2024 December 6 a new X-ray transient, designated EP241206a, was detected by WXT \citep{Yin2024} at 16:34:47 (UTC, $T_{0}$). The WXT observation of this event was split into two exposures as the transient exited the field of view during the satellite maneuver: the first exposure (ep11900018538wxt4s1) lasted for 895 s, and the second exposure (ep11908897056wxt16s1) lasted for 1074 s. 
The WXT data were reduced using standard procedures with the WXT Data Analysis Software (WXTDAS, v2.10; Y. Liu et al. 2026, in preparation) with the latest calibration database (CALDB; Cheng et al., in preparation).

\begin{figure}[t]
 \label{fig: spec_x}
 \centering
 \includegraphics[width = 0.7\columnwidth]{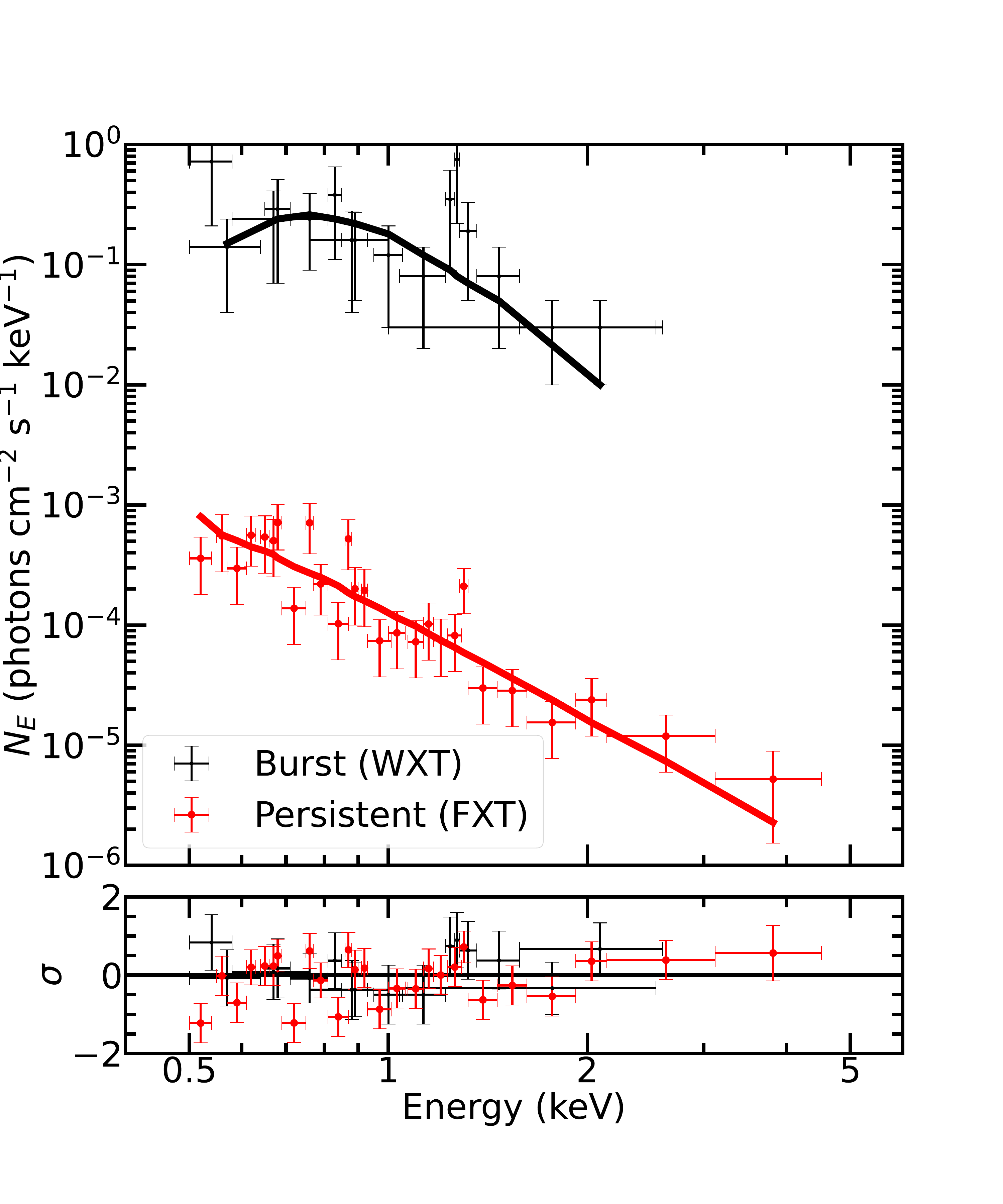}
 \caption{X-ray spectra of EP241206a observed with EP-WXT during the flare (black; $\Gamma = 2.0^{+0.5}_{-0.4}$) and with EP-FXT during the persistent state (red; $\Gamma = 2.94^{+0.16}_{-0.34}$). The top panel shows the spectra fitted with an absorbed power-law model, while the bottom panel displays the residuals of the fits.}
\end{figure}

The WXT position of EP241206a is RA = 02$^{\rm h}$18$^{\rm m}$48\fs48, Dec =+38\arcdeg54\arcmin50\farcs4 (J2000) with an uncertainty of 3.8 arcmin (90\% confidence level statistical and systematic). The average 0.5-4 keV spectrum can be fitted with an absorbed power-law with a photon index of $2.0^{+0.5}_{-0.4}$ (cstat/d.o.f. = 1.42), with a column density fixed at the Galactic value of $5.40 \times 10^{20}\  \rm{cm^{-2}}$ \citep{Wilms2000, HI4PI2016}. The average unabsorbed flux is $4.9^{+1.2}_{-1.2} \times 10^{-10}\ \rm{erg\ s^{-1}\ cm^{-2}}$ during the flare, corresponding to a fluence of $\sim2.5\times10^{-7}\ \rm{erg\ cm^{-2}}$ over the $\sim500$~s duration. The WXT spectrum is shown in Figure \ref{fig: spec_x}. 

\begin{figure}[t]
 \label{fig: ltcv}
 \centering
 \includegraphics[width = 0.45\textwidth]{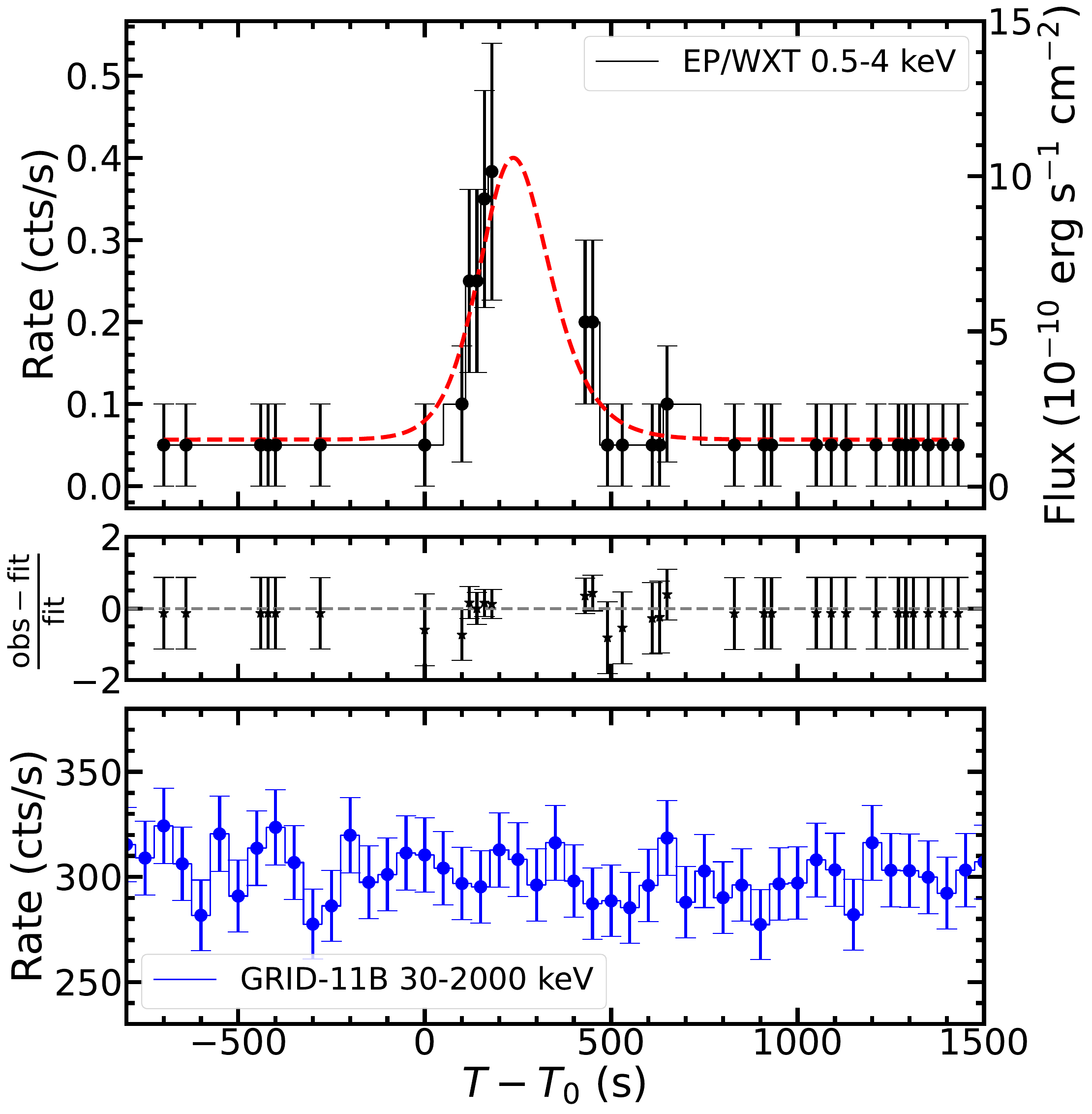}
 \caption{Light curves of EP241206a during the X-ray flare. The top panel shows the EP-WXT light curve, where black error bars represent the observed data and the red dashed line indicates the best-fit flare model. The middle panel shows the residuals of the light-curve fitting. The bottom panel presents the contemporaneous light curve measured with GRID-11B, with no flaring activity.}

\end{figure}

Figure \ref{fig: ltcv} presents the soft X-ray (0.5–4 keV) light curve recorded by WXT during the flare. Due to the source exiting and reentering the field of view during the observation, the WXT light curve is segmented into two parts, with a total observed duration of approximately 500 s. 

For comparison, the gamma-ray light curve measured with GRID-11B, the tenth satellite of the GRID Project \citep{Wen2019}, is also plotted; no significant variability was detected. A 95\% upper limit on the flux in the 30–2000 keV energy band was derived to be $8.2\times 10^{-8}\ \rm{erg\ s^{-1}\ cm^{-2}}$. During the event, \textit{Fermi} was Earth-occulted \citep{Meegan2009}, \textit{GECAM-C} was passing through the South Atlantic Anomaly \citep{Zhang2023}. The \textit{GECAM-B} and \textit{INTEGRAL} were not observing \citep{Li2021, vonKienlin2003}.

To characterize the temporal profile, we fitted the WXT light curve with a smooth pulse model proposed by \citet{Gowri2025}, which is constructed from two sigmoid-like (tanh) terms:
\begin{equation}
I(t)=\frac{A}{4}\left[1-\tanh\!\left(\frac{t-t_{\rm r}}{s_{\rm r}}\right)\right]\left[1+\tanh\!\left(\frac{t-t_{\rm l}}{s_{\rm l}}\right)\right] \, ,
\end{equation}
where $A$ is the normalization in units of count rate (counts\;s$^{-1}$), but does not necessarily equal the observed peak count rate because the pulse amplitude also depends on the degree of overlap between the rising and decaying sigmoid terms, $s_{\rm l}$ and $s_{\rm r}$ characterize the slopes of the rising and decaying edges, and $t_{\rm l}$ and $t_{\rm r}$ denote the half-transition times of the two sigmoid components, respectively \citep{Gowri2025}. The model provides an adequate fit to the light curve, with best-fit parameters $A = 11.0 \pm 1.3$ counts\,s$^{-1}$, $s_{\rm r} = 150 \pm 10$~s, $t_{\rm r} = 190 \pm 20$~s, $s_{\rm l} = 110 \pm 22$~s, and $t_{\rm l} = 230 \pm 15$~s. Given the time-averaged X-ray flux measured over the flare duration, we estimated the peak flux of the event as $\sim1.0\times10^{-9}\ \mathrm{erg\ s^{-1}\ cm^{-2}}$ by scaling the count rate.

Approximately 22 hr after the WXT trigger, a follow-up observation was conducted using FXT, starting at 14:34:29 UTC on 2024 December 7. The FXT data were processed using the FXT Data Analysis Software (FXTDAS, version 1.10), developed and maintained by the EP Science Center (EPSC), along with the latest FXT CALDB (version 1.1). Within the WXT error circle, an X-ray source was detected at RA = 02$^{\rm h}$18$^{\rm m}$47\fs76, Dec = +38\arcdeg54\arcmin49\farcs0 (J2000) with an uncertainty of 8\farcs1 (90\% C.L.\ statistical and systematic). 

The time-averaged FXT spectrum in 0.5-10 keV can be fitted by an absorbed power-law with a photon index of $2.94^{+0.16}_{-0.34}$ (cstat/d.o.f. = 1.10) and a column density fixed at the Galactic value of $5.40 \times 10^{20}\  \rm{cm^{-2}}$. The unabsorbed flux is $4.9^{+0.5}_{-0.4} \times 10^{-13}\ \rm{erg\ s^{-1}\ cm^{-2}}$ \citep{Wang2024}. The FXT spectrum is shown in Figure \ref{fig: spec_x}. This demonstrates a flux contrast of over 3 orders of magnitude between the flare and persistent states. 

ROSAT also detected this X-ray source during its all-sky surveys in the 1990s, cataloging it as 2RXS J021848.5+385423 \citep{Brinkmann1997, Boller2016}. The source position is RA = 02$^{\rm h}$18$^{\rm m}$48\fs12, Dec =+38\arcdeg54\arcmin52\farcs9 (J2000), with an uncertainty of approximately 1\arcmin, consistent with that of EP241206a. The historical unabsorbed flux in the 0.1–2.4 keV band is $4.7\times 10^{-13}\ \rm{erg\ s^{-1}\ cm^{-2}}$, consistent with the persistent level measured with FXT.

\begin{figure}
 \label{fig: flux_time}
 \centering
 \includegraphics[width = 0.45\textwidth]{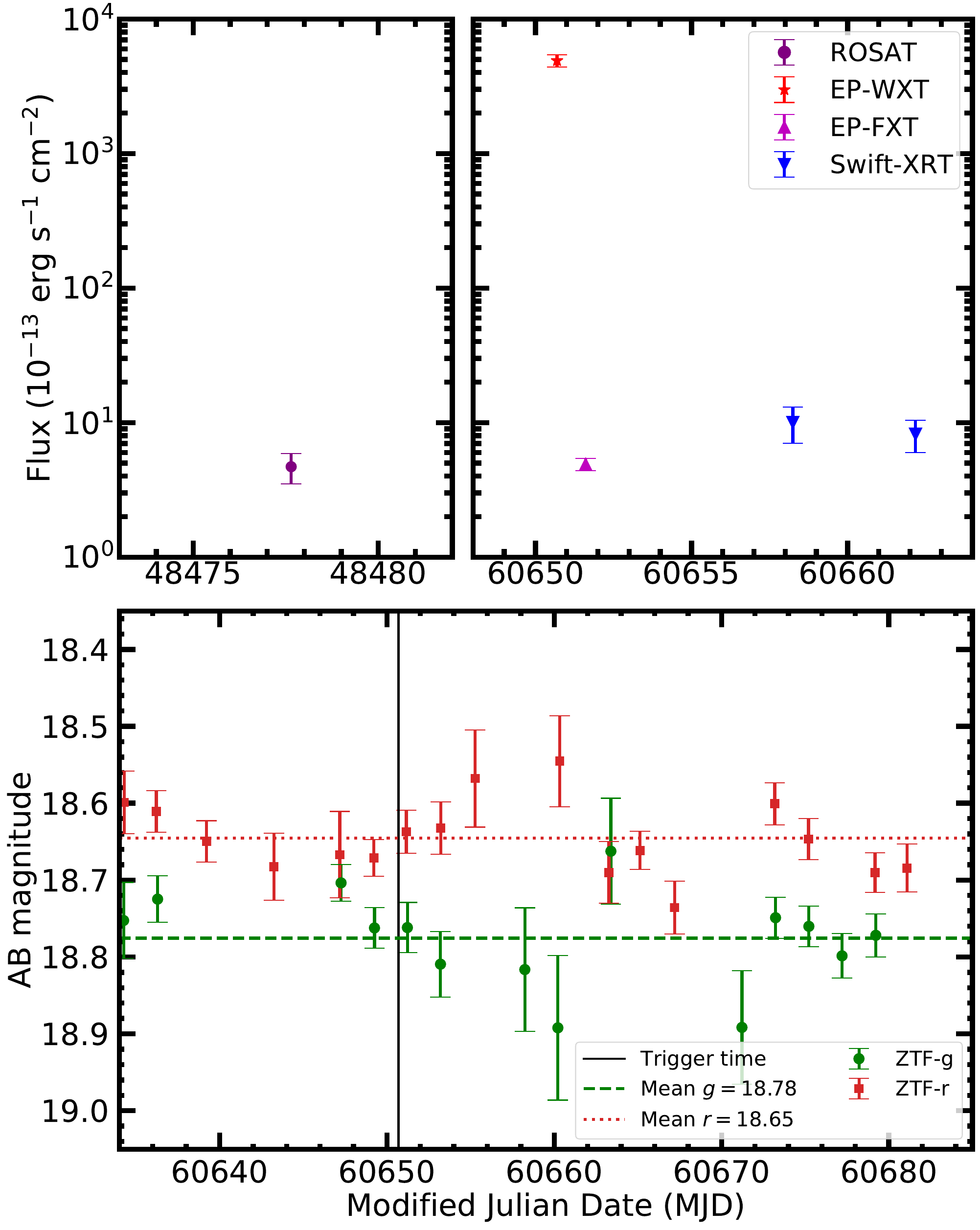}
\caption{Long-term light curve of EP241206a. The top panel shows the unabsorbed X-ray flux from ROSAT (purple), EP-WXT (red), EP-FXT (magenta), and Swift-XRT (blue). The bottom panel presents optical light curves from ZTF in the $g$ (green circles) and $r$ (red squares) bands. The vertical black solid line marks the trigger time of the X-ray flare. The horizontal green dashed and red dotted lines indicate the arithmetic mean magnitudes of the ZTF measurements displayed in the panel, $\langle g\rangle=18.78$ mag and $\langle r\rangle=18.65$ mag, respectively.}

\end{figure}

The Neil Gehrels Swift Observatory conducted two follow-up observations of the source with its X-Ray Telescope (XRT). The first observation (ObsID: 00018975001) was conducted on 2024 December 14 at 05:53:02 UTC, and the second (ObsID: 00018975002) on 2024 December 18 at 04:13:47 UTC \citep{Burrows2005}, following the trigger of EP241206a. A source was detected in both observations at a position RA = 02$^{\rm h}$18$^{\rm m}$48\fs00, Dec=+38\arcdeg54\arcmin51\farcs1 (J2000) with an uncertainty of 3\farcs1 (90\% confidence), consistent with the positions of the FXT source. The unabsorbed fluxes in the 0.3--10 keV band are $(1.0 \pm 0.3) \times 10^{-12}$ and $(8.2 \pm 2.2) \times 10^{-13}\ \rm{erg\ s^{-1}\ cm^{-2}}$ for the two observations using the XMM-Newton Upper Limit Server provided by ESA \citep{Saxton2022, Knig2022}, respectively. These values are consistent with the postflare flux measured with FXT and the historical flux with ROSAT. The long-term light curve is shown in Figure \ref{fig: flux_time}.

\subsection{Multiwavelength Counterparts and AGN Identification}

Within the FXT positional uncertainty, an optical counterpart was identified in the Pan-STARRS catalog (objID 154690346998737509) at RA = 02$^{\rm h}$18$^{\rm m}$47\fs97, Dec = +38\arcdeg54\arcmin50\farcs7, with a measured $g$-band magnitude of $g = 18.96 \pm 0.02$ mag \citep{Chambers2016, Brown2021}. 

The Zwicky Transient Facility (ZTF) covered the source region during the X-ray flare \citep{Masci2019}. 
The optical source has multiwavelength counterparts and has been classified as a candidate AGN \citep{Flesch2024}. Specifically, it is listed as MORX J021847.9+385450 in the Millions of Optical-Radio/X-ray Associations catalog (MORX), which compiles likely optical counterparts of radio and X-ray sources and provides source-type classifications \citep{Flesch2024}. This source is also associated with the radio source B3 0215+386. The ZTF $r$- and $g$-band light curves cover an extensive baseline that includes both the flaring and nonflaring intervals, yet they show no optical enhancement during the X-ray flare, shown in Figure \ref{fig: flux_time}. The plotted ZTF light curves were generated from the public ZTF data-release photometry; in ZTF DR24, the nonflaring mean magnitudes near the counterpart are $g\simeq18.78$ mag and $r\simeq18.65$ mag.

In the infrared band, the source was detected by the Wide-field Infrared Survey Explorer (WISE) and cataloged as WISE J021847.95+385450.8 \citep{Cutri2013, Schlafly2019}. The measured magnitudes are $W1 = 14.40 \pm 0.03$, $W2 = 13.25 \pm 0.03$, $W3 = 10.73 \pm 0.09$, and $W4 = 8.08 \pm 0.21$ mag. The resulting colors are $W1-W2=1.15\pm0.04$, $W2-W3=2.52\pm0.10$, and $W3-W4=2.65\pm0.23$ mag. The $W1-W2$ color places the source well within the WISE mid-infrared AGN selection region \citep{Assef2013}, while the four-band WISE colors are consistent with the three-dimensional WISE blazar locus in the $W1-W2$, $W2-W3$, and $W3-W4$ color space \citep{Abrusco2014}. This provides qualitative support for a jet-dominated AGN interpretation, although the WISE colors alone do not establish a blazar classification.

In the ultraviolet band, the source was detected by the Galaxy Evolution Explorer (GALEX) and is cataloged as GALEX J021847.9+385451 \citep{Bianchi2017}. The measured magnitudes are $NUV = 19.77 \pm 0.10$ mag and $FUV = 20.41 \pm 0.02$ mag. All the {\it GALEX} data used in this paper can be found in MAST \citep{10.17909/t9h59d}.

In the radio band, the source was first cataloged as B3 0215+386 in the Bologna 408 MHz B3 survey \citep{Ficarra1985}. The field has been observed with the Karl G. Jansky Very Large Array (VLA) on 1994 May 7 \citep{Laurent-Muehleisen1997}. A radio source was detected at RA = 02$^{\rm h}$18$^{\rm m}$48\fs13, Dec = +38\arcdeg54\arcmin52\farcs8 (J2000), with an uncertainty of 0\farcs5, located approximately 2\farcs7 offset from the optical counterpart of EP241206a. The reported flux density is $42 \pm 3$ mJy at 5~GHz. Given the offset from the optical nucleus, the radio emission may be associated with a large-scale jet \citep{Mullin2008, Godfrey2013, Hardcastle2020}.

The source was also detected with the VLA Sky Survey (VLASS; \citealt{Lacy2020, Gordon2021}) in 2018, with a measured flux density of $65.31 \pm 0.13$ mJy at 3~GHz. More recently, the field was observed with the Australian SKA Pathfinder (ASKAP; \citealt{Hotan2021}). In the RACS-High data release \citep{Duchesne2025}, the source was detected at 1.65~GHz during 2021--2022 with a flux density of $116.0 \pm 0.5$ mJy.

\begin{figure}
 \centering
 \includegraphics[width = 0.45\textwidth]{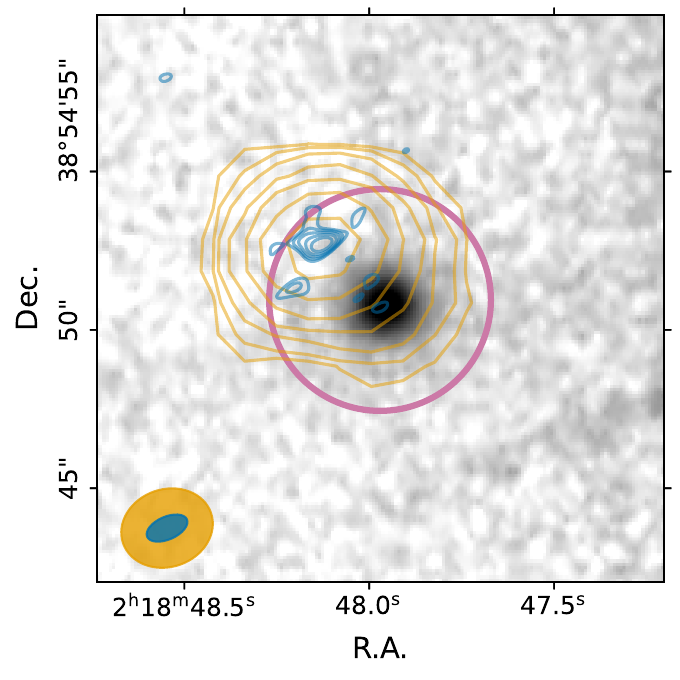}
 \caption{Multiwavelength images around EP241206a. The background image is from Pan-STARRS $i$-band. The purple circle indicates the 90\% Swift-XRT position uncertainty. Radio contours (4, 8, 16, 32, 64, 128, and 256 times the rms noise level) from VLASS (orange) and (4, 8, 16, 32, and 64 times the rms noise level) VLA (blue) are overlaid. The rms noise levels are $0.17\ \rm{mJy\ beam^{-1}}$ (VLASS) and $0.33\ \rm{mJy\ beam^{-1}}$ (VLA), respectively.}
 \label{fig: image}
\end{figure}

We reprocessed the archival VLA data (project: AK360, PI: R.I. Kollgaard) following standard procedures using the Common Astronomy Software Applications \citep[CASA;][version 6.4.1]{CASATeam2022}. The data were calibrated using J1331+305 (3C286) as the flux and bandpass calibrator, and J0205+322 as the phase calibrator. Imaging was performed with the {\tt tclean} task using Briggs weighting with a ROBUST parameter of 0. The measured flux density is consistent with the published value. Figure \ref{fig: image} shows Pan-STARRS $i$-band optical image around the source, overlaid with radio contours from VLA and VLASS. 

\begin{figure*}
 \centering
 \includegraphics[width = 0.9\textwidth]{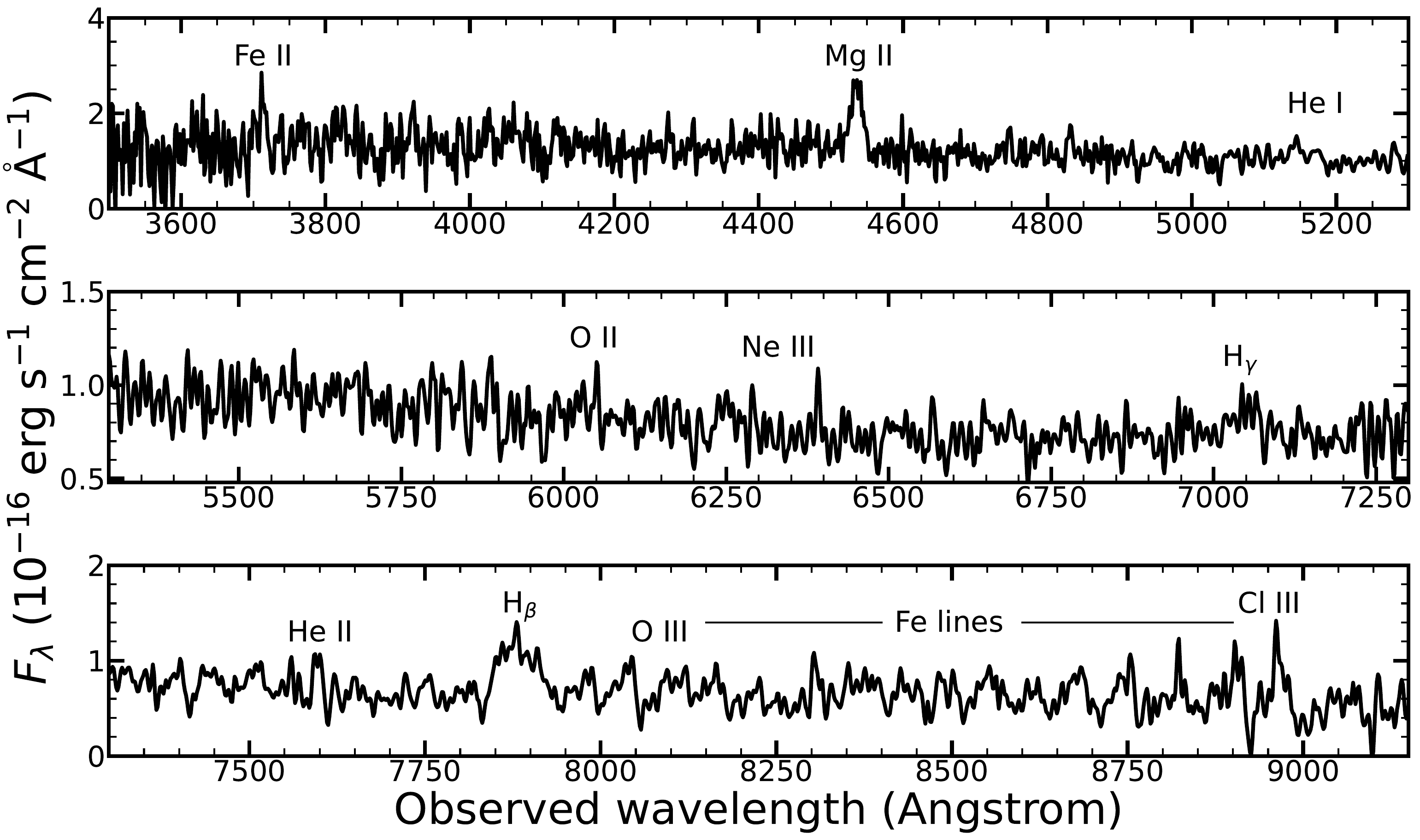}
 \caption{Optical spectrum of EP241206a obtained with LCO. Prominent emission lines are marked.}
 \label{fig: opspec}
\end{figure*}

We requested spectroscopic observations of the Pan-STARRS source with Las Cumbres Observatory (LCO) 2 m telescope equipped with FLOYDS~\citep[through the Global Supernova Project;][]{Brown2013LCO}. The observation was performed on 2024 December 19 at 08:40:04 UTC. The spectrum is displayed in Figure \ref{fig: opspec}. Several prominent broad emission lines are identified, including \MgII\ and \hbeta, characteristic of an AGN. We determined a redshift of $z = 0.62$ based on the line wavelengths.

\section{Discussion}\label{Discu}
\subsection{Association Probability}\label{asso_pro}

Prior to discussing the physical origin of EP241206a, we first discuss the association between the X-ray transient and the AGN counterpart. The significance of the association depends strongly on which counterpart class is considered. We therefore estimate the chance-coincidence probability for several pre-defined catalog classes.

We adopt the 90\% WXT localization radius of $r_{\rm WXT}=3\farcm8$, corresponding to an error-circle area of $A_{\rm WXT}=\pi r_{\rm WXT}^2=1.26\times10^{-2}\ {\rm deg^2}$. For each catalog class, we estimate the single-trial probability as $P_1=1-\exp(-\Sigma A_{\rm WXT})$, where $\Sigma$ is the sky surface density of the selected catalog sources within the relevant survey footprint. For the 18 WXT transients with averaged flux comparable to or exceeding that of EP241206a detected up to the time of this event, we also quote the trial-corrected probability $P_{18}=1-(1-P_1)^{18}$.

We considered two counterpart classes. First, we used ROSAT-detected quasars/AGNs associated with 2RXS sources and required their X-ray fluxes to be no fainter than the persistent ROSAT counterpart of EP241206a \citep{Boller2016}. Second, as the closest available multiband comparison sample, we used radio- and X-ray-emitting AGNs from the RASS--Green Bank (RGB) catalog, additionally requiring a ROSAT X-ray flux comparable to or brighter than that of EP241206a \citep{Laurent-Muehleisen1997, Brinkmann1997}.

For the flux-limited ROSAT-detected AGN class, we obtain $P_1\simeq4.2\times10^{-4}$ and $P_{18}\simeq7.5\times10^{-3}$. For the flux-limited RGB class, the corresponding probabilities are $P_1\simeq2.7\times10^{-4}$ and $P_{18}\simeq4.9\times10^{-3}$.

Because the counterpart is detected in both X-rays and radio, we use the 2RXS AGN and RGB samples as physically motivated comparison classes, with the radio requirement motivated by the possible role of a relativistic jet in producing the rapid flare. We do not use the follow-up optical/infrared properties, such as the WISE colors or broad emission lines, to refine the probabilities because this would be an a posteriori selection; the resulting values are therefore conditional chance-coincidence estimates that make the association plausible but do not establish it.

\subsection{Physical Origins of the X-Ray Flare}

\subsubsection{AGN Properties}

\begin{figure}[t]
 \label{fig: em_line}
 \centering
 \includegraphics[width = 0.45\textwidth]{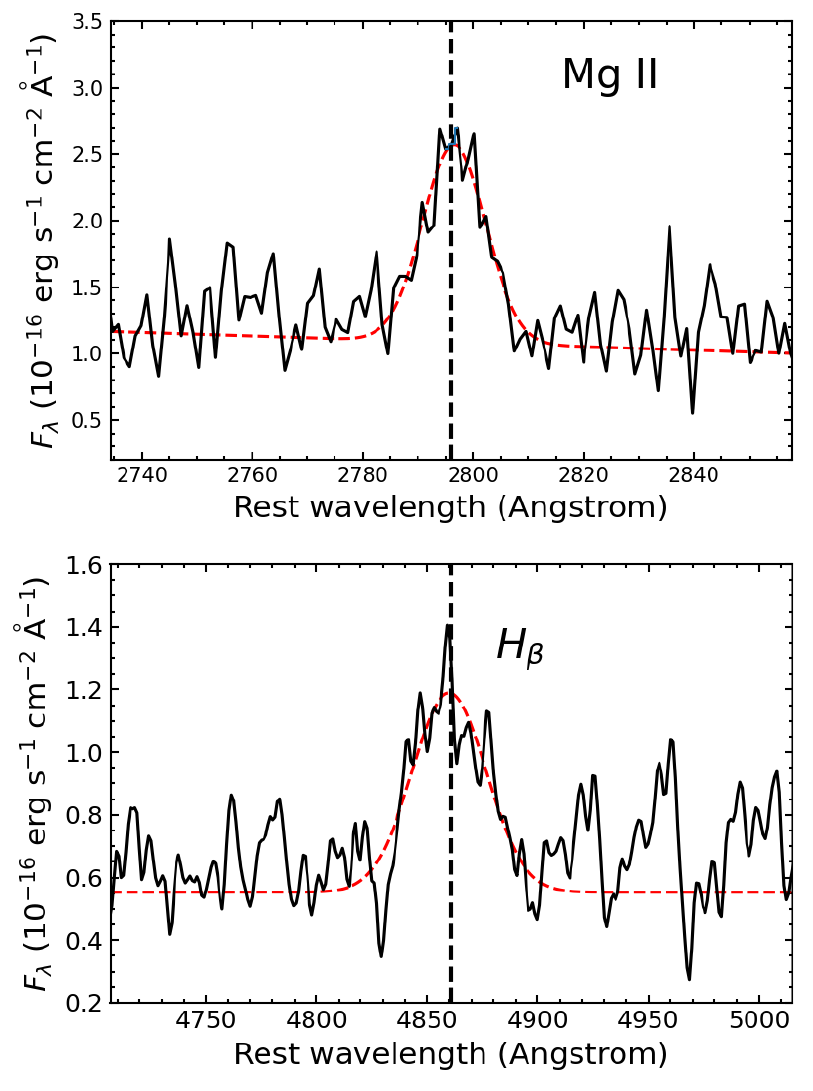}
 \caption{Optical emission-line fitting for EP241206a. The top panel shows the \MgII\ line, and the bottom panel shows the \hbeta\ line, together with the best-fit models.}
\end{figure}

The optical spectrum shows prominent broad \MgII\ and \hbeta\ emission lines. We fitted the emission features with a Gaussian plus a linear continuum, using the \texttt{Spectrum} and \texttt{WaveCoord} tools from the \texttt{mpdaf.obj} module of MPDAF \citep{MPDAF2017}, and obtained FWHM(\MgII)\,$= 11.4$\,\AA\ ($\sim1220\ \mathrm{km\ s^{-1}}$) and FWHM(\hbeta) = 30.5\,\AA\ ($\sim1880\ \mathrm{km\ s^{-1}}$) (see Figure \ref{fig: em_line}). These values have been corrected for instrumental broadening based on the spectral resolution of the LCO spectrograph \citep{Shporer2010, Pickles2014}.

Using the broad \hbeta\ and the continuum luminosity at the rest 5100~\AA, $L_{5100}\simeq1.8\times10^{44}\ {\rm erg\ s^{-1}}$, we estimated the black hole mass through the standard single-epoch virial relation \citep{Vestergaard2009} 

\begin{align}
\log\left(\frac{M_{\rm BH}}{M_\odot}\right)
= 0.91 + 0.5\log\left(\frac{L_{5100}}
{10^{44}\ {\rm erg\ s^{-1}}}\right)
\nonumber \\
+ 2\log\left(\frac{{\rm FWHM}_{\rm H\beta}}
{\rm km\ s^{-1}}\right) \, .
\label{eq:mbh}
\end{align}

and obtained 
$M_{\rm BH} = 3.8^{+0.7}_{-0.8}\times10^{7}\,M_\odot$. While such estimates carry systematic uncertainties associated with the geometry and kinematics of the broad-line region, the derived mass provides a useful order-of-magnitude constraint for assessing the characteristic size and energetics of the central engine.

We assume a luminosity distance of 3.69\,Gpc to the source given the redshift and standard cosmology ($h = 0.7$, $\Omega_{\rm m}=0.3$, and $\Omega_\Lambda=0.7$; \citealt{Wright2006}). The FXT spectrum yields a rest-frame 2--10~keV intrinsic luminosity of $L_{2-10}\simeq2.3\times10^{44}\ \mathrm{erg\ s^{-1}}$. The burst-averaged rest-frame 2--10~keV isotropic-equivalent luminosity of the flare is $L_{2-10,\rm avg}\simeq7.2\times10^{47}\ \mathrm{erg\ s^{-1}}$. Assuming no appreciable spectral evolution, the peak-to-average WXT count-rate ratio of $\sim2$ gives an estimated peak luminosity of $L_{2-10,\rm pk}\simeq1.4\times10^{48}\ \mathrm{erg\ s^{-1}}$. The isotropic-equivalent X-ray energy integrated over the flare is $E_{\rm iso,X}\sim3\times10^{50}\ \mathrm{erg}$. From the luminosity-dependent relationship provided in \citealt{Duras2020}, we adopt a bolometric correction $k_X\equiv L_{\rm bol}/L_{2-10}=16$. This gives a persistent-state bolometric luminosity of $L_{\rm bol}\sim3.7\times10^{45}\ \mathrm{erg\ s^{-1}}$, an apparent burst-averaged bolometric luminosity of $\sim1.1\times10^{49}\ \mathrm{erg\ s^{-1}}$, and an estimated apparent peak bolometric luminosity of $\sim2.2\times10^{49}\ \mathrm{erg\ s^{-1}}$. Given the black hole mass estimated from the broad \hbeta, the corresponding Eddington ratios are about 0.8 in the persistent state, $\sim2.3\times10^3$ averaged over the flare, and $\sim4.6\times10^3$ at peak.

\subsubsection{X-Ray Flare Due to AGN Jet Activity}

The X-ray flare (EP241206a) is characterized by an unusually short duration of $\sim$500~s (or $\sim$300\,s in the rest frame) and an extreme luminosity increase. Given the estimated black hole mass of $M_{\rm BH}=3.8\times10^{7}\,M_\odot$, 
the minimum dynamical timescale at the innermost orbit (e.g., $6GM/c^2$ for a Schwarzschild black hole) is 18.6\,ks, and the light-crossing time at the same radius is 1.2\,ks. Thus, a duration of 300\,s, together with a beaming-dependent burst-averaged Eddington ratio of $\sim2.3\times10^3$, suggests that the flare does not originate from the bulk accretion flow, but is instead associated with a relativistic jet subject to Doppler boosting \citep[e.g.,][]{Urry1995, Hovatta2009, Blandford2019, Raiteri2025}.

A natural realization is a localized dissipation episode in a magnetically dominated inner jet. Relativistic magnetic reconnection can produce compact plasmoids that move relativistically within the bulk jet and generate rapid flares \citep{Giannios2009, Giannios2010, Sironi2014, Sironi2016, Petropoulou2016}. This mechanism has been applied to minute-scale AGN variabilities; for example, \citet{Shukla2020} derived a plasmoid flare timescale of $t_{\rm p}\sim493$~s for 3C~279, suggesting that reconnection-powered plasmoids can produce flares as short as the one observed in EP241206a.

For a compact emitting region of characteristic size $R_{\rm em}$, the observed variability timescale is given by the standard light-crossing argument corrected for Doppler compression and cosmological time dilation \citep[e.g.,][]{Tavecchio1998, Hovatta2009, Petropoulou2016}:
\begin{equation}
    t_{\rm obs} \simeq (1+z)\frac{R_{\rm em}}{c\,\delta_{\rm eff}},
\end{equation}
where $\delta_{\rm eff}$ is the effective Doppler factor of the emitting region. Estimating the expected variability timescale therefore requires plausible ranges for both $R_{\rm em}$ and $\delta_{\rm eff}$.

The plasmoid size can be estimated from the characteristic length $\ell^{\prime}$ of the reconnection layer in the jet comoving frame. Here $w_{\rm f}$ denotes the final plasmoid width when it leaves the layer. Particle-in-cell simulations show that plasmoids form self-consistently in such layers \citep{Sironi2016}. In the model of \citet{Petropoulou2016}, small and fast plasmoids have $w_{\rm f}\simeq0.04\ell^{\prime}$, while large and slow ones have $w_{\rm f}\simeq0.2\ell^{\prime}$.

Since \citet{Shukla2020} inferred a reconnection-region scale of order $10^2R_g$ for a minute-scale AGN flare, these relations imply plasmoid widths from a few to tens of $R_g$. Identifying $R_{\rm em}$ with $w_{\rm f}$ to order unity therefore gives $R_{\rm em}$ from a few to tens of $R_g$.

Variability and VLBI studies of radio-loud AGNs commonly infer jet Doppler factors from a few to several tens \citep{Hovatta2009}. In reconnection models, the effective Doppler factor can be further enhanced by the motion of the plasmoid within the bulk jet \citep{Petropoulou2016}. Thus, a plausible range for $\delta_{\rm eff}$ is also from a few to tens, including both the bulk jet motion and the additional plasmoid motion.

For $R_{\rm em} / r_{\rm g}$ and $\delta_{\rm eff}$ ranging from a few to tens, this gives $t_{\rm obs}\sim30$--$3000$~s, encompassing the observed duration of EP241206a, $\sim500$~s.

As an order-of-magnitude estimate, the reconnecting magnetic power is $P_{\rm rec}\sim u_B v_{\rm rec} A$, where $u_B=B^2/8\pi$, $v_{\rm rec}=\beta_{\rm rec}c$, and $A\sim2\pi R_{\rm em}^2$ is the inflow area of a compact reconnection region \citep[e.g.,][]{Sironi2014, Sironi2016, Petropoulou2016}. For emission approximately isotropic in the plasmoid frame, the observed bolometric luminosity is boosted as $L_{\rm obs}=\delta_{\rm eff}^4L'$; equivalently, the band luminosity scales as $\delta_{\rm eff}^{3+\alpha}$, which is close to $\delta_{\rm eff}^4$ given the observed X-ray spectral index $\alpha_X = \Gamma-1\simeq1$ \citep[e.g.,][]{Urry1995, Petropoulou2016}. If a fraction $\eta_X$ of the dissipated magnetic power is radiated in the observed X-ray band,
\begin{equation}
    L_{X,\rm obs}
    \sim \delta_{\rm eff}^{4}\eta_X
    \frac{B^2 c R_{\rm em}^2 \beta_{\rm rec}}{4} \; ,
\end{equation}

where $B$ is the magnetic-field strength in the reconnecting region and $\beta_{\rm rec}$ is the reconnection inflow speed in units of $c$. We define
$\eta_X \equiv f_{\rm rec}\epsilon_{\rm rad}f_X$ as the effective fraction of the reconnecting magnetic power that emerges in the observed X-ray band, where
$f_{\rm rec}$ is the fraction of the dissipated magnetic energy transferred to radiating particles, $\epsilon_{\rm rad}$ is their radiative efficiency, and $f_X$ is the fraction of the radiated power falling within the observed X-ray band. Relativistic reconnection simulations generally find inflow speeds of order $\beta_{\rm rec}\sim0.1$ \citep{Sironi2014,Sironi2016}.

In the plasmoid-flare model of \citet{Petropoulou2016}, the adopted plasmoid growth rates are $\beta_g\simeq0.06$--0.1, while the energy-transfer fraction is $f_{\rm rec}\simeq0.15$--0.5. The remaining factors entering $\eta_X$ depend on the particle distribution, radiative cooling, and broadband spectral energy distribution and are not directly constrained by the available observations of EP241206a. We therefore adopt $\eta_X=0.1$ as an illustrative normalization while retaining its explicit dependence in the luminosity estimate. With parameters appropriate for a compact inner-jet region, the luminosity becomes

\begin{align}
    L_{X,\rm obs}
    &\sim 7\times10^{47}
    \left(\frac{\delta_{\rm eff}}{10}\right)^4
    \left(\frac{\eta_X}{0.1}\right)
    \left(\frac{\beta_{\rm rec}}{0.1}\right)
    \left(\frac{B}{10^4\ {\rm G}}\right)^2
    \nonumber\\
    &\quad \times
    \left(\frac{R_{\rm em}}{17R_g}\right)^2
    \left(\frac{M_{\rm BH}}{3.8\times10^7M_\odot}\right)^2
    {\rm erg\ s^{-1}} .
\end{align}

This estimate is comparable to the apparent peak luminosity of EP241206a, $L_{X,\rm pk}\simeq1.4\times10^{48}\ {\rm erg\ s^{-1}}$. The fiducial magnetic field is high but remains within the range expected near a compact AGN jet base. For example, \citet{Baczko2016} inferred magnetic fields of $\sim10^3$--$10^4\ {\rm G}$ near the base of the jet in NGC~1052, and similarly strong fields are expected in magnetically powered jet-launching models \citep{Blandford2019}.

For the adopted parameters, the synchrotron cooling time of the X-ray-emitting electrons is much shorter than the magnetic reconnection timescale. Sustaining the X-ray emission throughout the flare therefore requires continued particle acceleration, as can occur during an ongoing magnetic-reconnection episode. The spectral behavior is also qualitatively consistent with a reconnection-powered plasmoid flare, although it is not by itself a decisive diagnostic. \citet{Petropoulou2016} show that small relativistic plasmoids can produce short high-energy synchrotron flares, with emission extending to the UV/soft-X-ray band and flux-doubling times as short as $\sim0.22$~hr in their representative model. This supports an X-ray flare origin qualitatively. However, the observed photon index of EP241206a depends on the detailed particle distribution, magnetic field, cooling history, and possible inverse-Compton contribution, and thus a dedicated broadband spectral-energy-distribution model would be required before using the X-ray slope as a quantitative constraint.

The lack of a gamma-ray flare does not rule out this scenario. If the event was produced by a localized plasmoid whose high-energy emission was temporarily beamed toward the observer, the contrast can be much larger in X-rays than in the optical band, where the persistent AGN and host-galaxy emission may dilute a weak transient component. The gamma-ray constraint is also not decisive because the relevant high-energy coverage was incomplete: the contemporaneous GRID-11B limit is useful, but \textit{Fermi}/GBM was Earth-occulted during the WXT flare, and the expected MeV--GeV flux depends sensitively on the poorly constrained SSC or external-Compton component. Thus, the absence of a reported gamma-ray counterpart should be interpreted as a weak constraint rather than as evidence against the jet scenario.

An alternative, more extreme realization of this picture is the ``jet-in-the-jet'' scenario \citep{Giannios2009, Giannios2010}, in which magnetic reconnection produces plasmoids that are themselves moving relativistically within the bulk jet flow, resulting in an additional Doppler boost on top of that of the jet. This mechanism can further enhance the apparent luminosity and compress the variability timescale and is therefore also capable of accounting for the rapid and luminous flare observed in EP241206a.

However, this scenario implies a population of minijets with a range of orientations and Doppler factors, so more frequent lower-amplitude flares would also be expected. In EP241206a, only one prominent flare is detected, with no clear evidence for such lower-amplitude, short-timescale variability. Moreover, if the persistent emission is jet-dominated, this interpretation would imply numerous bright flares, which is difficult to reconcile with the source's modest long-term variability. These considerations suggest that although the jet-in-the-jet scenario remains viable, it may require some fine-tuning or a rare, short-lived favorable beaming configuration to account for the observed properties of EP241206a.

\subsubsection{GRBs in the AGN Disk}

An alternative is that the event corresponds to a long gamma-ray burst (GRB). The observed X-ray duration ($\sim$500~s) and apparent peak luminosity are broadly consistent with those of long GRBs \citep{Zhang2018}, and the relatively soft X-ray spectrum is also not uncommon in this class \citep{Sun2025, Li2025, Dai2026}. In this interpretation, the temporal properties of EP241206a would not be unusual, but rather typical of a long GRB observed in X-rays.

In the context of a GRB occurring in an AGN environment, the observed properties of the burst may differ from those of standard GRBs \citep{Perna2021, Zhu2021}. The dense and radiatively active medium in the nuclear region, such as an AGN disk, can strongly affect both the prompt emission and the afterglow. Numerical studies suggest that jets propagating in such environments may experience enhanced interaction with the surrounding medium, leading to softer observed spectra, suppressed high-energy emission, and modified temporal profiles \citep{Lazzati2023}. In particular, the afterglow emission can be significantly weakened or smeared out due to radiative diffusion and the bright background emission from the AGN itself \citep{Wang2022, Ray2023}.

In this scenario, several observational features of EP241206a can be naturally accommodated. The GRB optical emission, or afterglow, could be outshone by the persistent AGN and host-galaxy emission. The nondetection in the gamma-ray band is also compatible with a soft, X-ray-dominated event, as expected if the emission is reprocessed or intrinsically shifted to lower energies. For a power-law spectrum with photon index $\sim2$, the expected flux in the 30--1000~keV band would be $\sim2\times10^{-9}\ \rm{erg\ s^{-1}\ cm^{-2}}$, well below the sensitivity of GRID-11B \citep{Wen2019}.

In fact, EP has recently detected several short, GRB-like transients without gamma-ray emission \citep{ZhangYQ2025}. EP240305a may represent a GRB-like transient given radio signatures of a relativistic jet (R. Ma et al. 2026, in preparation); EP240414a may involve strong interaction of a choked jet with dense environment \citep{Sun2025, Bright2025, vanDalen2025}; EP241021a could be explained by moderately off-axis GRB \citep{Shu2025, Busmann2025}; EP241113a is consistent with a baryon-loaded outflow \citep{Dai2026}; EP250108a can be interpreted as an intrinsically soft X-ray flash \citep{Li2025}. This diversity suggests that gamma-ray faint GRB-like transients detected by EP do not originate from a uniform class. In the case of EP241206a, the lack of afterglow signatures makes it difficult to further constrain its physical origin.

\begin{figure}[t]
 \label{fig: amati}
 \centering
 \includegraphics[width = 0.45\textwidth]{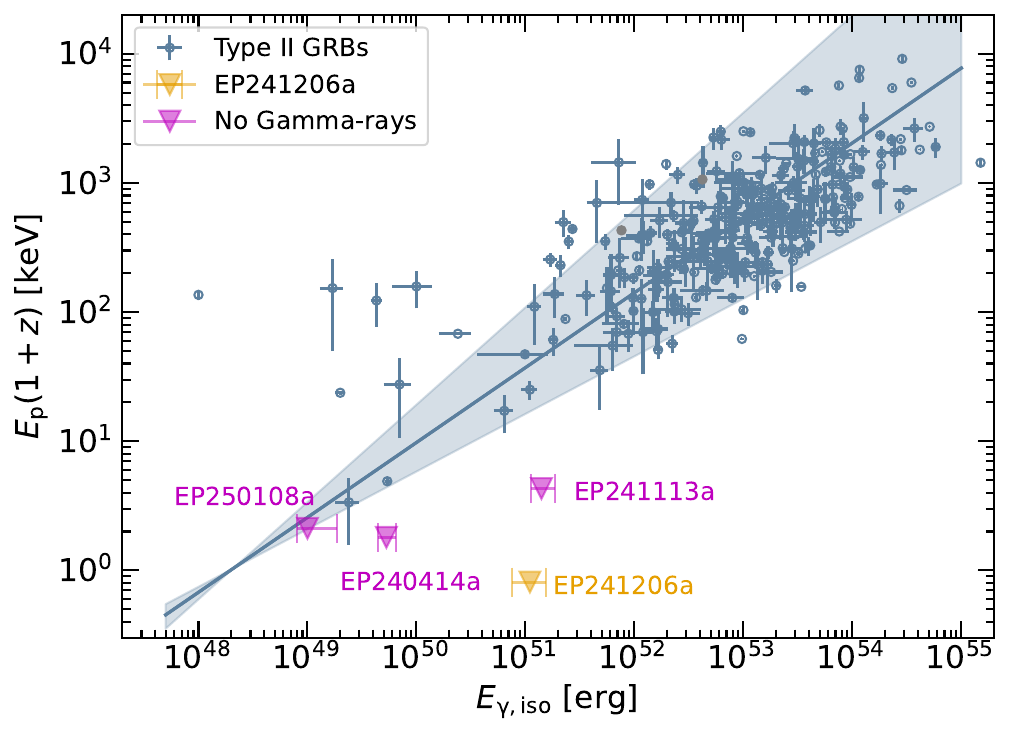}
 \caption{Amati relation between the rest-frame peak energy $E_{\rm p}(1+z)$ and the isotropic-equivalent energy $E_{\gamma,\rm iso}$. Blue points represent Type II GRBs, with the solid line and shaded region indicating the best-fit relation and its 1$\sigma$ dispersion. Purple upper-limits denote three soft GRB candidates detected by EP without gamma-ray emission. The orange upper-limit marks EP241206a.}
\end{figure}

However, if we assume that the event is a long GRB with a single power-law spectrum of photon index $\sim 2$, we obtain $E_{\rm iso} = 1.1_{-0.3}^{+0.4}\times10^{51}\ \rm erg$ \citep{Zhang2018}. As shown in Figure \ref{fig: amati}, the source does not lie on the Amati relation for Type II GRBs \citep{Amati2002, Dai2026}. In addition, the association of a long GRB with an AGN has not been firmly established observationally. Therefore, although the GRB scenario provides a simple explanation for several observed properties, these considerations suggest that it should be regarded as a plausible alternative rather than a definitive interpretation.

\subsection{Implications for Fast X-Ray Transients}

Both the persistent (AGN emission) and flaring (EP241206a) states of the source, with a contrast over 3 orders of magnitude, can be detected with today's major X-ray telescopes. If the source were located at a distance 10 times farther (at a redshift of $\sim$4), the optical magnitude would be about 24, making spectroscopic identification with ground-based optical facilities extremely challenging. 
The X-ray flux of the AGN emission would decrease to $\sim5\times10^{-15}\ \mathrm{erg\ s^{-1}\ cm^{-2}}$, approaching the detection limit of Chandra \citep{Evans2024} and XMM-Newton \citep{Hasinger2000, Brunner2007} with a moderate exposure time. However, the peak of the flare, with a flux of $\sim1.5\times10^{-11}\ \mathrm{erg\ s^{-1}\ cm^{-2}}$, is still detectable with WXT \citep{Zhang2022}. 

EP241206a may therefore represent a favorably oriented example of a broader population of jet-dominated AGN flares whose persistent emission is typically invisible in X-rays or optical, potentially contributing to the population of unidentified or hostless X-ray transients detected with WXT, such as EP240408a \citep{Zhang2025}.

\section{Conclusion}\label{conclusion}

EP241206a is a fast X-ray transient discovered with EP-WXT, displaying a single-peaked soft X-ray flare with a duration of $\sim500$~s and a peak 0.5--4~keV flux of $\sim1.0\times10^{-9}\ \mathrm{erg\ s^{-1}\ cm^{-2}}$ ($L_{2-10,{\rm pk}}\simeq1.4\times10^{48}\ {\rm erg\ s^{-1}}$ if associated with the AGN).  It is spatially coincident with an AGN at a redshift of $z=0.62$. The persistent X-ray emission of the AGN measured with ROSAT, Swift and EP-FXT is around $\sim5\times10^{-13}\ \mathrm{erg\ s^{-1}\ cm^{-2}}$. The broad \hbeta\ from the AGN indicates a central black hole mass of $M_{\rm BH}\sim3.8\times10^{7}\ M_\odot$, suggesting an Eddington ratio of 0.8 at the persistent state.

However, the physical association is not firmly established: the 18-trial chance probability is large for broad AGN/radio priors, but decreases to $\sim0.5$--$0.8\%$ for flux-limited ROSAT or RGB counterpart classes. A chance alignment therefore cannot be ruled out. If the X-ray flare and AGN are physically associated, its peak-to-persistent flux contrast exceeds three orders of magnitude and its apparent peak Eddington ratio reaches $\sim4.6\times10^3$. The flare may be due to three different physical origins: magnetic reconnection in the inner regions of a relativistic jet that produces compact, rapidly moving emitting regions whose emission is strongly Doppler-boosted, which naturally explains the short timescale and high apparent luminosity with a compact plasmoid; a jet-in-the-jet model, which can provide additional Doppler boosting but may require favorable orientation and flare-rate fine tuning; or a GRB in the AGN disk, which can accommodate a soft X-ray-dominated event but remains less constrained because no afterglow or decisive gamma-ray counterpart was detected.

EP241206a may represent a favorably oriented example of a broader population of unidentified fast X-ray transients detected with WXT.

\begingroup
\let\internallinenumbers\relax
\let\endinternallinenumbers\relax
\begin{acknowledgments}
H.F. acknowledges funding support from the Strategic Priority Research Program of the Chinese Academy of Sciences, and China's Space Origins Exploration Program. This work makes use of observations from the Las Cumbres Observatory network. The LCO team is supported by NSF grants AST-2308113 and AST-1911151. RS acknowledges hospitality at the IHEP during part of this work.
\end{acknowledgments}
\endgroup

\bibliography{ms.bib}
\bibliographystyle{aasjournal}

\end{document}